\documentclass{sig-alternate-05-2015}
\usepackage{multirow}
\usepackage{booktabs}
\usepackage{subcaption}
\usepackage{graphicx}
\usepackage{subfig}
\usepackage{tablefootnote}
\usepackage{color}
\usepackage{mdwlist}
\usepackage[dvipsnames]{xcolor}
\usepackage{soul}

\newcommand{\hide}[1]{}

\begin{document}

\setcopyright{acmcopyright}

\doi{10.475/123_4}

\isbn{123-4567-24-567/08/06}

\conferenceinfo{KDD '16}{August 13--17, 2016, San Francisco, CA, USA}

\acmPrice{\$15.00}

%
\conferenceinfo{KDD}{'16 San Francisco, California USA}
\CopyrightYear{2016} 

\title{
Firebird: Predicting Fire Risk and \\Prioritizing Fire Inspections in Atlanta
}


\numberofauthors{8} 
\author{
%
%
\alignauthor
Michael Madaio\\ 
       \affaddr{Carnegie Mellon University}\\
       \affaddr{Pittsburgh, PA, USA}\\
       \email{mmadaio@cs.cmu.edu}
\alignauthor Shang-Tse Chen\\
        \affaddr{Georgia Tech}\\
        \affaddr{Atlanta, GA, USA}\\
        \email{schen351@gatech.edu}
\alignauthor Oliver L. Haimson\\ 
       \affaddr{University of California, Irvine}\\
       \affaddr{Irvine, CA, USA}\\
       \email{ohaimson@uci.edu}
\and       
\alignauthor Wenwen Zhang\\ 
       \affaddr{Georgia Tech}\\
       \affaddr{Atlanta, GA, USA}\\
       \email{wzhang300@gatech.edu}
\alignauthor Xiang Cheng\\
        \affaddr{Emory University}\\
       \affaddr{Atlanta, GA, USA}\\
       \email{xcheng7@emory.edu}
\alignauthor Matthew Hinds-Aldrich\\
       \affaddr{Atlanta Fire Rescue Dept.}\\
       \affaddr{Atlanta, GA, USA}\\
       \email{mhinds-aldrich@atlantaga.gov}
\and  
\alignauthor Duen Horng Chau\\
        \affaddr{Georgia Tech}\\
       \affaddr{Atlanta, GA, USA}\\
      \email{polo@gatech.edu}
\alignauthor Bistra Dilkina\\
        \affaddr{Georgia Tech}\\
       \affaddr{Atlanta, GA, USA}\\
      \email{bdilkina@cc.gatech.edu}
}

\maketitle
\begin{abstract}

The Atlanta Fire Rescue Department (AFRD), like many municipal fire departments, actively works to reduce fire risk by inspecting commercial properties for potential hazards and fire code violations. 
However, AFRD's fire inspection practices relied on tradition and intuition, with no existing data-driven process for prioritizing fire inspections or identifying new properties requiring inspection. 
In collaboration with AFRD, we developed the \textit{Firebird}  framework 
to help municipal fire departments identify and prioritize commercial property fire inspections, using machine learning, geocoding, and information visualization. 
Firebird computes fire risk scores for over 5,000 buildings in the city, with true positive rates of up to 71\% in predicting fires. 
It has identified 6,096 new potential commercial properties to inspect, based on AFRD's criteria for inspection.
Furthermore,
through an interactive map, Firebird integrates and visualizes fire incidents, property information and risk scores to help AFRD make informed decisions about fire inspections. 
Firebird has already begun to make positive impact at both local and national levels.
It is improving AFRD's inspection processes and Atlanta residents' safety, and was highlighted by National Fire Protection Association (NFPA) as a best practice for using data to inform fire inspections.
\end{abstract}



\begin{figure}[!t]
\centering
    \includegraphics[width=0.45\textwidth]{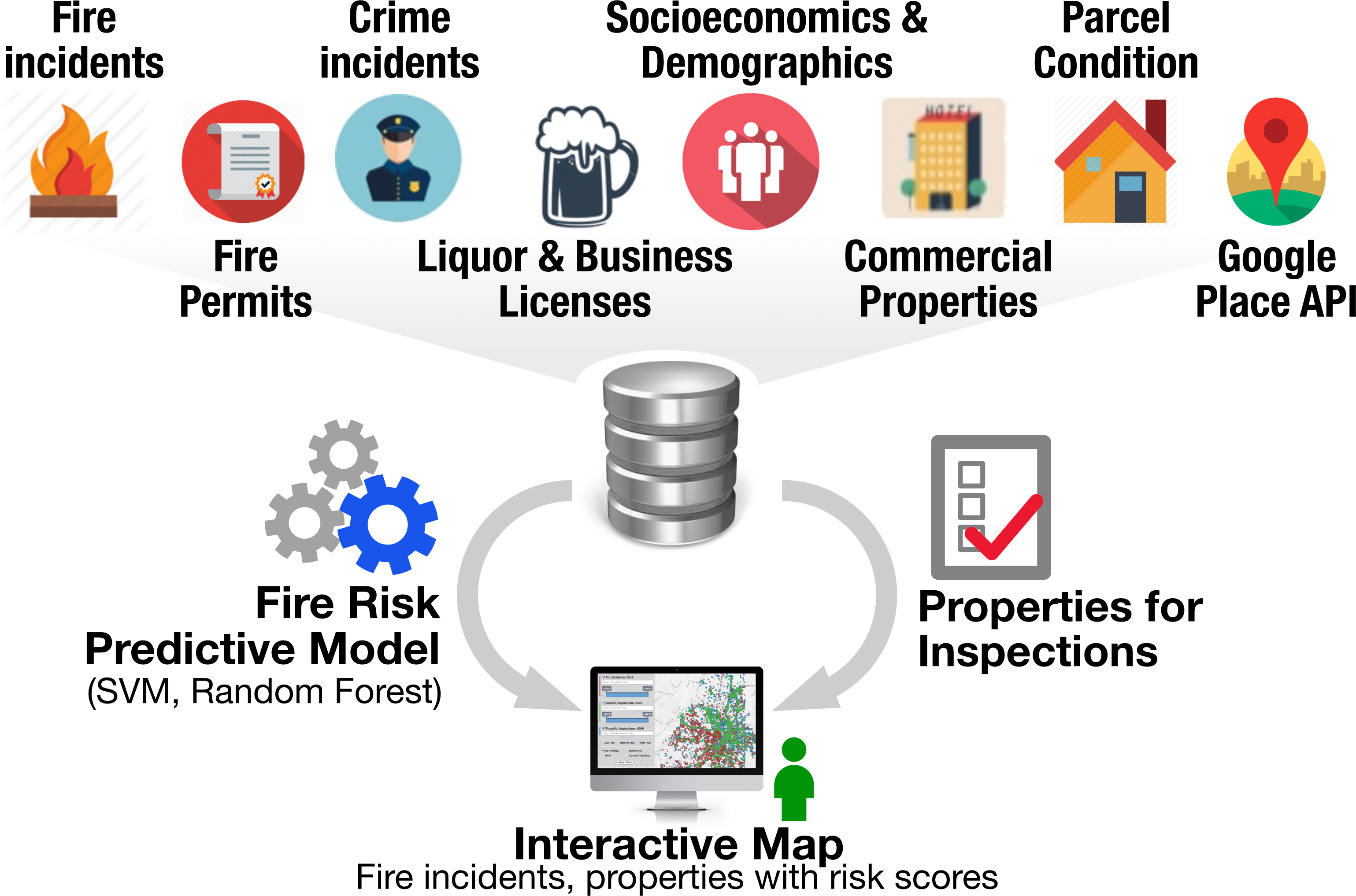}
\caption{\textbf{Firebird Framework Overview.} By combining 8 datasets, Firebird identifies new commercial properties for fire inspections.
Its fire risk predictive models (SVM, random forest) and interactive map help AFRD prioritize fire inspections and personnel allocation.}
\label{fig:crownjewel}
\end{figure}

\section{Introduction}

In 2014 alone, there were 494,000 structure fires in the United States, causing 2,800 civilian deaths and \$9.8 billion in property damage \cite{nfpa}. Municipal fire departments, as the Authority Having Jurisdiction (AHJ), are responsible for enforcing applicable fire codes to reduce the risk of structure fires. The City of Atlanta Fire Rescue Department (AFRD), like many other fire departments, conducts regular commercial property inspections to ensure that they comply with the city's Code of Ordinances \cite{firecode} for fire prevention and safety. With an annual average of nearly 650 structure fires and 2,573 annual commercial inspections, the AFRD Community Risk Reduction Section wanted to both identify uninspected properties and ensure that the properties being inspected were those at greatest risk of fire. 
Through a partnership between the City of Atlanta and the Data Science for Social Good (Atlanta) program, 
our research team developed the \textit{Firebird} framework for identifying and prioritizing property fire inspections, based on  fire department criteria and historical fire risk,
tackling two important challenges:

\textbf{Challenge 1: Property Identification.} The AFRD Community Risk Reduction Section knew that the 2,573 annually inspected commercial properties were not all of the commercial properties in the city of Atlanta, but they did not have a way to obtain a more complete list of commercial properties that potentially needed inspection. The existing process for AFRD's property inspections involved a legacy system of paper file records and inspections conducted on the basis of pre-existing permits, without a robust process for identification, selection, and prioritization of new properties to inspect. In addition, the variety of data sources AFRD had compiled to inform their inspections were inconsistent, incomplete, and were often at different levels of granularity. Thus, cleaning and merging the datasets to identify which inspectable properties in the city had fallen through the cracks required significant effort. By integrating data from a variety of government and commercial sources, we discovered 19,397 potential new commercial properties to inspect, based on the property usage types that the Atlanta Code of Ordinances specifies require inspection. 

\textbf{Challenge 2: Fire Risk Prediction.} Because 19,397 new commercial property inspections is far greater than the current number of annual commercial property inspections, and far more than AFRD's current staff of fire inspectors can reasonably inspect, we developed a method to prioritize those inspections based on their fire risk. 
First, we created a joined dataset of building- and parcel-level information variables, for 8,223 commercial properties\footnote{We will be referring to \emph{buildings} and \emph{properties} as two distinct concepts throughout this paper. The AFRD conducts property inspections and issues permits to the owners of those ``inspectable spaces," which are properties. However, it is the physical structure of buildings that catch fire, and thus, when we built  predictive models, we did so with information about the buildings themselves. This is significant because one property may contain multiple buildings, while another building may contain multiple properties.}.
Then, we built predictive models of fire risk using machine learning approaches, including Support Vector Machine (SVM)~\cite{cortes1995support} and Random Forest~\cite{breiman2001random}.
These models achieve true positive rates (TPRs) of up to 71.36\% (in predicting fires) at a false positive rate (FPR) of 20\%.
As our most important goal is to save lives, a higher TPR outweighs the increase in FPR.
The resulting fire risk scores were then assigned to over 5,000 commercial properties to help ARFD prioritize inspections.

\textbf{Contributions \& Impact.} With Firebird, AFRD can now use data about historical fires to inform their fire inspections and more efficiently utilize their inspection personnel capacity. 
The challenges that Firebird addresses are not unique to AFRD or the City of Atlanta; 
many municipal agencies across the country work to integrate a variety of data sources to inform decision-making at all levels of governance. 
Specifically, many fire safety departments are seeking effective prioritization of property inspections and allocation of inspection resources, given limited inspection personnel and large numbers of inspectable properties.
Firebird has already begun to improve AFRD's inspection processes. Its major contributions include:
\begin{itemize*}  \itemsep1pt \parskip0pt \parsep0pt

\item \textbf{Discovering new properties.} 
Firebird improves the safety of Atlanta residents and visitors by identifying 19,397 previously unidentified inspectable commercial properties.

\item \textbf{Predictive fire risk model.} 
Firebird correctly predicts more than 70\% of commercial fires (at 20\% FPR), and applies the resulting fire risk scores to over 5,000 properties 
 to help ARFD prioritize inspections.

\item \textbf{Impact to Atlanta: Firebird at work.} 
Through an interactive map, Firebird integrates and visualizes fire incidents, property information and  inspections, and risk scores to inform the decision-making processes of AFRD fire inspectors, executive staff, and their Community Risk Reduction Section for inspection prioritization and inspection personnel allocation.

\item \textbf{National impact: reusable end-to-end framework for inspection prioritization.} 
Firebird provides an explicated model for other municipalities and agencies to use to identify new properties and prioritize commercial property inspections based on fire risk. 
This project was highlighted by the \textit{National Fire Protection Association} (NFPA) at the \textit{Smart Enforcement Workshop} for fire service professionals across North America as a best practice for using data to inform fire inspections.
\end{itemize*}

\begin{figure*}[!ht]
\centering
    \includegraphics[width=0.85\textwidth]{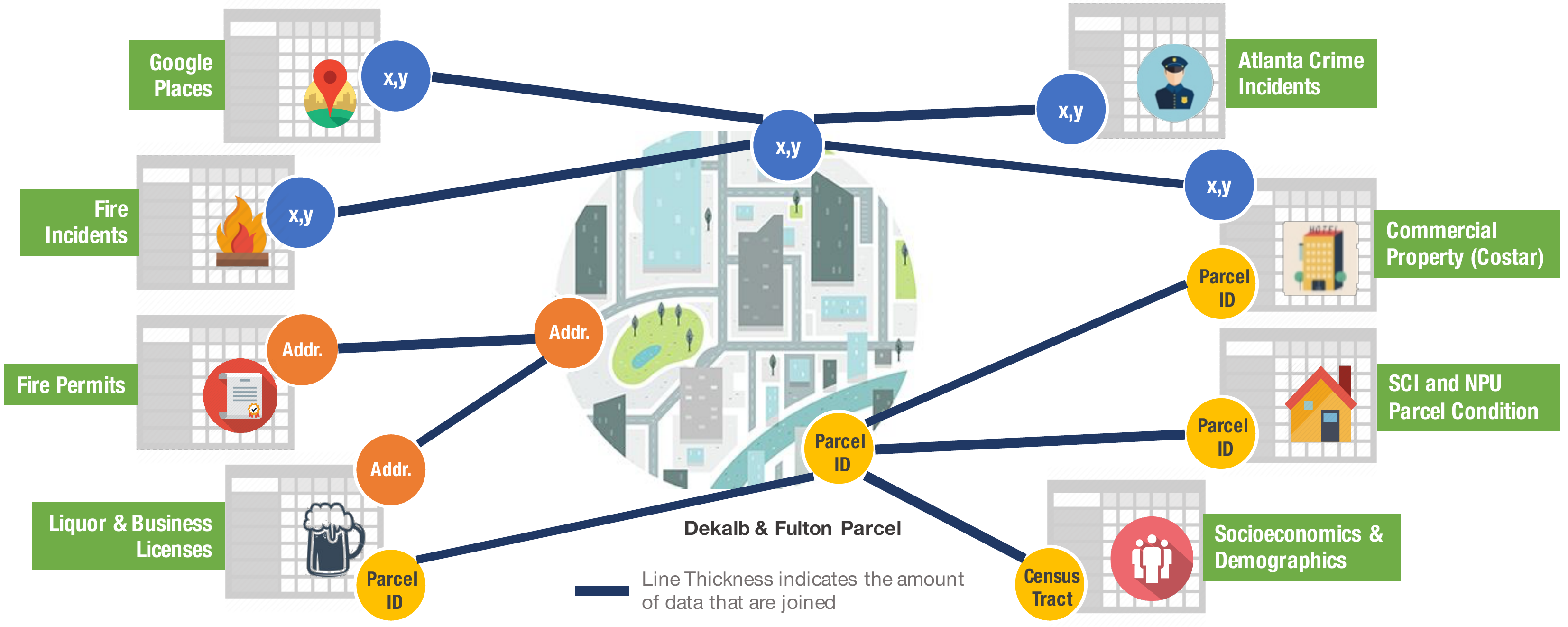}
\caption{Joining eight datasets using three spatial information types (geocode, address, parcel ID).}
\label{fig:joining}
\end{figure*}

\section{Related Work}
Risk prediction models have been widely used in many domains, including health care~\cite{kansagara2011risk}, student performance evaluation~\cite{lakkaraju2015machine}, and accounting fraud detection~\cite{mcglohon2009snare}. However, urban fire risk prediction has received relatively less attention, despite its obvious importance.

\textbf{Forest fire prediction.} Much of the prior work on data-driven fire risk prediction has targeted woodland and forest fires, such as in Italy \cite{lapucci05}, Greece \cite{iliadis05},  and Portugal \cite{de01}. They used different methods, such as neural networks~\cite{de01}, fuzzy algebra~\cite{iliadis05}, and decision trees \cite{lapucci05} to support the allocation of firefighting, fire prevention, and foliage recuperation resources to the areas of highest fire risk. The features they used, such as soil type and topography, are very different from the ones typically used in urban fire prediction like construction material and property usage type.

\textbf{Community-level urban fire prediction.} Prior work in data-driven urban fire risk prediction tends to work at the region or community level~\cite{clare12, dacosta15}, rather than the property- or building-level, which is the unit that the Atlanta fire inspectors are assigned to inspect. For instance, \cite{clare12} undertook a randomized controlled trial of community fire risk education efforts, targeting high-risk residential communities. However, their method for identifying the high-risk areas was to create a point-distribution map of residential structure fires and draw ellipses to capture the areas of densest concentration of fire incidents. A more methodologically rigorous approach, as seen in \cite{dacosta15}'s work on optimizing smoke-alarm inspections, joins data from the American Community Survey and American Housing Survey to predict municipal blocks most likely to have homes without functioning smoke alarms, using a Random Forest. Our work similarly uses publicly available datasets to predict properties most likely to be in need of inspection, but differs in that we offer a fire risk prediction score for individual commercial properties, rather than municipal residential blocks.

\textbf{Property-level urban fire prediction.} There is limited work on predicting fire risk at the property or building level. In British Columbia, \cite{garis14} developed a risk-based model for determining the frequency of commercial property fire inspections, using static and dynamic building-level characteristics. They scored each property by its level of compliance on prior inspections and by a set of risk metric components such as building classification, age, and presence of sprinklers. However, as they acknowledge, the weights and selection of those components were based on their fire code, and not on historical data on features that were highly predictive of fire, such as we utilize in our work. 

The nearest precedent for our research with AFRD is the recent work from the New York Mayor's Office of Data Analytics (MODA) with the Fire Department of New York (FDNY) to build a ``Risk-Based Inspection System'' (RBIS) \cite{copeland15}. They built a data-driven model to identify structures at greatest fire risk, to better prioritize FDNY's inspection process, using a set of structural and behavioral information about those properties. However, due to a lack of detailed information on their technical approach, it is unclear how it may apply to AFRD's scenario.

In both the FDNY RBIS initiative and our work with AFRD, a key challenge emerged: the difficulty of joining disparate sets of data about commercial properties, gathered from various city departments without a shared convention for building ID numbers, consistent address formats, or strict internal quality control practices to ensure the datasets are accurate and up-to-date. We differ from \cite{copeland15} and \cite{garis14} by providing a clear method for identifying new inspectable commercial properties that the fire department is not already aware of. Further, our work goes beyond \cite{copeland15} by presenting here a detailed comparison of the performance of several machine learning algorithms for predicting the fire risk of commercial properties, and by incorporating them into an interactive GIS visualization for use by the AFRD fire inspectors and Community Risk Reduction Section, following \cite{dobrica12}. 

\section{Data Description}
An essential step before identifying and prioritizing potential properties to inspect is to join the data about commercial properties from multiple sources. This was done to construct as complete a picture as possible for the properties in Atlanta needing inspection, as required by the Atlanta Code of Ordinances. 
After the data joining, we identified 19,397 new potential commercial properties to inspect, through a process of property discovery that utilized AFRD and City of Atlanta fire code criteria. See Table~\ref{tab:inspections} for a summary of the different lists of total commercial property inspections and commercial buildings we will be referring to throughout this paper.

\begin{table}[!h]
\sffamily
  \centering

    \begin{tabular}{lll}
    \toprule
    \textbf{Name} & \textbf{Count} \\
    \midrule
    \multicolumn{1}{l}{\textbf{Current} annual inspections} & \multicolumn{1}{r}{2,573}\\
    \multicolumn{1}{l}{\textbf{Long list} of potential new inspections\protect\footnotemark} &
    \multicolumn{1}{r}{19,397}\\
    \multicolumn{1}{l}{\textbf{Short list} of potential new inspections} & \multicolumn{1}{r}{6,096}\\
    \multicolumn{1}{l}{\textbf{Current} $+$ \textbf{short list} inspections} & \multicolumn{1}{r}{8,669}\\
    \multicolumn{1}{l}{\textbf{Current} $+$ \textbf{short list} inspections with risk score} & \multicolumn{1}{r}{5,022}\\
    \multicolumn{1}{l}{Properties for building predictive model} & \multicolumn{1}{r}{8,223}\\
    \bottomrule
    \end{tabular}%
    
  \caption{Summary of inspection and building lists}
  \label{tab:inspections}%
\end{table}%

\begin{table*}[htbp]
\small
\sffamily
  \centering
    \begin{tabular*}{0.9\textwidth}{lll}
    \toprule
    \textbf{Source} & \textbf{Name}  & \textbf{Description} \\
    \midrule
    \multicolumn{1}{l}{\multirow{2}[0]{*}{Atlanta Fire Rescue Department}} & \multicolumn{1}{l}{Fire Incidents} & \multicolumn{1}{l}{Fire incidents from 2011 - 2015} \\
\cmidrule{2-3}
    \multicolumn{1}{l}{} & \multicolumn{1}{l}{Fire Permits} & \multicolumn{1}{l}{All permits filed by AFRD in 2012-2015} \\
\midrule
    \multicolumn{1}{l}{\multirow{3}[0]{*}{City of Atlanta}} & \multicolumn{1}{l}{Parcel} & \multicolumn{1}{l}{Basic information for each parcel in Atlanta} \\
\cmidrule{2-3}
    \multicolumn{1}{l}{} & \multicolumn{1}{l}{Strategic Community Investigation} & \multicolumn{1}{l}{Information regarding parcel conditions} \\
\cmidrule{2-3}
    \multicolumn{1}{l}{} & \multicolumn{1}{l}{Business Licenses} & \multicolumn{1}{l}{All the business licenses issued in Atlanta} \\
\midrule
    \multicolumn{1}{l}{\multirow{2}[0]{*}{Atlanta Police Department}} & \multicolumn{1}{l}{Crime} & \multicolumn{1}{l}{2014 crime in Atlanta} \\
\cmidrule{2-3}
    \multicolumn{1}{l}{} & \multicolumn{1}{l}{Liquor Licenses} & \multicolumn{1}{l}{All filed liquor licenses by Police Department} \\
\midrule
    \multicolumn{1}{l}{Atlanta Regional Commission} & \multicolumn{1}{l}{Neighborhood Planning Unit} & \multicolumn{1}{l}{Boundary data for each Atlanta neighborhood} \\
\midrule
    \multicolumn{1}{l}{\multirow{2}[0]{*}{U.S. Census Bureau}} & \multicolumn{1}{l}{Demographic} & \multicolumn{1}{l}{Household number, population by race and age} \\
\cmidrule[0.01em]{2-3}
    \multicolumn{1}{l}{} & \multicolumn{1}{l}{Socioeconomic} & \multicolumn{1}{l}{Household median income} \\
\midrule
    \multicolumn{1}{l}{CoStar Group, Inc} & \multicolumn{1}{l}{CoStar Properties} & \multicolumn{1}{l}{Commercial property information} \\
\midrule
    \multicolumn{1}{l}{Google Place APIs} & \multicolumn{1}{l}{Google Place} & \multicolumn{1}{l}{Information regarding places from Google Maps} \\
    \bottomrule
    \end{tabular*}%
  \caption{Data Sources Summary}
  \label{tab:datasummary}%
\end{table*}%

\footnotetext{We provided AFRD with two lists of potential properties: one longer list that was the most extensive that we could provide, and another shorter list that was more manageable to display on a map, refined using the most frequently inspected property usage types.}

\subsection{Data Sources}
Firebird uses data from multiple sources, as tabulated in Table~\ref{tab:datasummary}. AFRD provided us with a dataset of $2,543$ historical fire incidents from July 2011 to March 2015, among which $34.3\%$ were commercial fires. The dataset includes information about fire incidents, such as time, location, type and cause of fire. AFRD also provided a dataset of fire inspections, containing $32,488$ inspection permit records from 2012 to 2015. The inspection data includes information such as inspected property types, address, and time of inspections. We also obtained structural information about commercial properties from a dataset purchased by AFRD from the CoStar Group, a commercial real estate agency. This dataset includes building-level features such as year built, building material, number of floors and units, building condition and other information. A total of 8,223 commercial properties are documented by the CoStar Group in the City of Atlanta. 

While CoStar offers building-level information, parcel data from Atlanta's Office of Buildings provides parcel-level information, such as property value, square footage, address, and other information about each parcel (a unit of land surrounding building(s)). The business license dataset obtained from the City of Atlanta's Office of Revenue provides information about businesses that own commercial properties. The business licenses dataset has 20,020 records with over 20 features including business type, business name, address, owner, etc. For non-business commercial properties (e.g., schools, churches, daycare centers), we obtained such data from Google Places API and State of Georgia Government.

To offer more information about properties for building a predictive risk model, we also obtained socioeconomic and demographic data from the U.S. Census Bureau, liquor license and 2014 crime data from the Atlanta Police Department, and Certificate of Occupancy (CO) data from the Atlanta Office of Buildings. All of these data sources contributed to discovering new inspections and developing our predictive model for commercial fire risk estimation. 

\subsection{Data Joining}
A critical step of this study was to join different datasets together so that data from different sources about the same building or property could be unified to create the most complete picture of a given property. For instance, by joining fire incident and commercial property data together, we can obtain a general idea regarding which commercial properties caught fire in the past five years. Furthermore, by joining commercial property data with data from the commercial real estate reports like the CoStar Group or the SCI Report, we can generate a more comprehensive view regarding specific characteristics of buildings, such as the structure and parcel condition, and even vacancy information.  

We joined the datasets together based primarily on spatial location information. There are three types of spatial or location information in our datasets: longitude and latitude, address information, and the parcel identification number, which is a unique ID number created by Fulton and DeKalb county\footnote{The City of Atlanta is comprised of two separate counties, Fulton and Dekalb. 
Although both county governments provided building information,  their parcel ID numbering schemes were not consistent.
Thus, building information had to joined using addresses and coordinates.} for tax purposes. We then performed a location join based on the above three types of location information. The variety of spatial information types, and our method for joining them is illustrated in Figure~\ref{fig:joining}. One obstacle we encountered was that spatial information had different formatting standards across the datasets. For example, the addresses from the CoStar Group were all in lowercase, with road names abbreviated instead of fully spelled out, while datasets from the multiple departments of the City of Atlanta tend to use a more consistent address format. Therefore, a spatial information cleaning process was conducted before joining the datasets directly. The address location information from different datasets was first validated using Google Geocoding API. The API can auto-correct some misspellings of address information. After validation, addresses were then reformatted using US Postal Service's address validation API. The coordinate information was processed in ESRI ArcGIS software to filter out data points falling outside of the City of Atlanta. The cleaned datasets were then joined together based on the formatted addresses from the USPS API and the coordinate information from ArcGIS.

\section{Identifying New Properties \\Needing Inspection}
To discover new properties, we first needed to understand what types of properties currently required fire inspections by AFRD, and we then identified other similar properties. In the current fire inspection permit dataset, we found more than 100 unique occupancy usage types, such as restaurants, motor vehicle repair facilities, textile storage, schools, children's day care centers, etc. To identify other similar commercial properties, we joined the list of currently inspected properties with the Atlanta Business License data by matching both the spatial location (identified through the joining process explained in Section 3.2) and the business name. 

We discovered that, in addition to the 2,573 currently inspected properties, there were approximately 19,397 properties of the same occupancy usage types as the city's current inspections. For instance, the Fire Code of Ordinances stipulates that motor vehicle repair facilities require inspection, due to the presence of flammable or combustible materials. However, only 186 of a total of 507 of those facilities in the city were on the list of current annual property inspections, suggesting that many or all of 321 remaining facilities should be inspected. However, because some occupancy types, such as ``miscellaneous business service," may have many properties that are not actually required for inspection, we created a shorter, more refined list of 6,096 new potential property inspections (instead of the 19,397 mentioned above), including only the top 100 most frequently inspected property usage types. 
We discovered these properties from a variety of data sources, including the Atlanta Department of Revenue's Business License dataset, the liquor license dataset from the Atlanta Police Department, the Georgia Department of Education's child care and preschool database, and Google Places API. 
We used the Google Places API to supplement the other datasets primarily because it provided more up-to-date information about some of our most commonly inspected property types, such as restaurants, bars, nightclubs, schools, churches, gas stations, etc, and because it proved especially useful for discovering properties that required inspection, but were not in the Business License dataset as they did not belong to any ``business" category (e.g., churches). 
Google Places API served as a ``bridge" to cross-check properties from different datasets, increasing the accuracy of our property discovery process.

In the process of identifying new properties needing inspection, the most challenging part was 
to determine how buildings with different names (or IDs, or address formats) in various datasets may actually refer to the same building. We had to ensure that the properties on our new inspectable property list were unique and were not already on the list of currently inspected properties, after the aforementioned datasets were joined together. Different approaches were attempted to ensure the uniqueness and novelty of the properties on our potential list. The most reliable and efficient method was found to be joining different datasets in pairs using geocoding and approximate (``fuzzy'') string matching to approximately match both the business name and the address. We used Google Maps Geocoding API for geocoding and a Python library~\cite{strmat} to match the strings based on the edit distance. From the joined dataset, a final property list was extracted that contained information from all the available data sources.

\section{Predictive Model of Fire Risk}
However, 19,397 new properties (or even the shorter list of 6,096) is far more than AFRD is able to add to their annual property inspections, and moreover, not all of those properties are likely to need inspection at the same level of priority. We therefore created a predictive model to generate a fire risk score based on the building- and parcel-level characteristics of properties that had fire incidents in the last five years. This model was built using the scikit-learn machine learning package in Python~\cite{scikit-learn}. The model uses 58 independent variables to predict fire as an outcome variable for each property. 

\subsection{Data Cleaning}
After joining various datasets together to obtain building- and parcel-level information, significant data cleaning still needed to occur. The bulk of the data cleaning process involved finding the extent of the missing data and deciding how to deal with that missingness. Our missingness procedures were designed to minimize deletion of properties with missing data, because a significant number of the properties in our model had NA values (not available) for many variables (such as the structure condition of a building, which is only known if the building was inspected by the CoStar Group before). For each property with missing data for a particular feature, we replaced missing values with 0 when appropriate. We also included a binary feature indicating whether each property had missing data for each feature. We used log transformation for variables with a large numerical range, such as the ``for sale" price of properties.

\subsection{Feature Selection}
After merging the datasets, we had a total of 252 variables for each property. We manually examined each variable to determine whether it may be relevant to fire prediction, and excluded many obviously non-predictive variables in this initial process (such as the phone number of the property owner, or property ID numbers). 
We then used forward and backward feature selection processes to determine each variable's contribution to the model, and removed the variables that did not contribute to a higher predictive accuracy. Our final model includes only 58 variables. We then expanded categorical variables into binary features. For example, the zip code variable was expanded into 37 binary features, and for each property only one zip code was coded as $1$ (all zip codes were designated as $0$ if the property's zip code data was missing). After the expansion, we had 1127 features in total.

\subsection{Evaluation of the Models}
\label{sec:evaluate}
We chose to validate our model using a time-partitioned approach. A fire risk model would ideally be tested in practice by predicting which properties would have a fire incident in the following year, and then waiting a year to verify which properties actually did catch fire. Because we wanted to effectively evaluate the accuracy of our model without waiting a year to collect data on new fires, we simulated this approach by using data from fire incidents in July 2011 to March 2014 as training data to predict fires in the last year of our data, April 2014 to March 2015.

We used grid search with 10-fold cross validation on the training dataset to select the best models and parameters.
The models we tried included \textit{Logistic Regression}~\cite{nelder1972generalized}, \textit{Gradient Boosting}~\cite{friedman2002stochastic}, \textit{Support Vector Machine} (SVM)~\cite{cortes1995support}, and \textit{Random Forest}~\cite{breiman2001random}. 
SVM and Random Forest performed the best,
with comparable performances (see Table~\ref{table:split}). For SVM, we used RBF kernel with $C=0.5$ and $\gamma = \frac{10}{\#\text{features}}$. For Random Forest, we used 200 trees with a maximum depth of 10 for each tree.

We then trained SVM and Random Forest on the whole training set using the best parameters and generated predictions on the testing set. Note that the training and testing sets consist of the same set of properties, but the different labels correspond to fires in a different period of times. This is a valid approach because we didn't use any information that we would only know after the training period, i.e., fires in 2015.

\begin{figure}[!t]
    \centering
     \begin{subfigure}[b]{0.32\textwidth}
        \includegraphics[width=\textwidth]{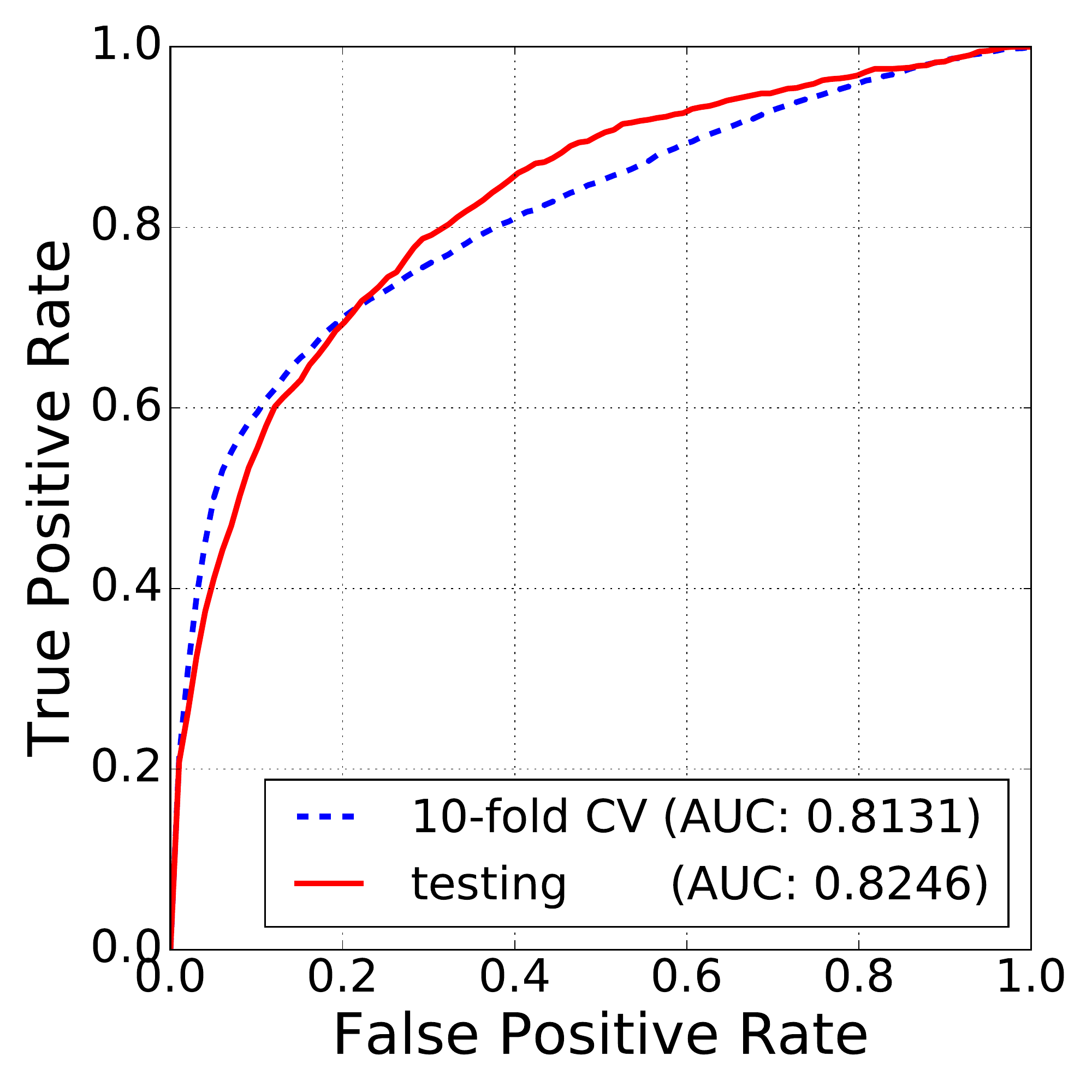}
        \caption{Random Forest}
         \label{fig:rf_curve}
     \end{subfigure}
    \begin{subfigure}[b]{0.32\textwidth}
        \includegraphics[width=\textwidth]{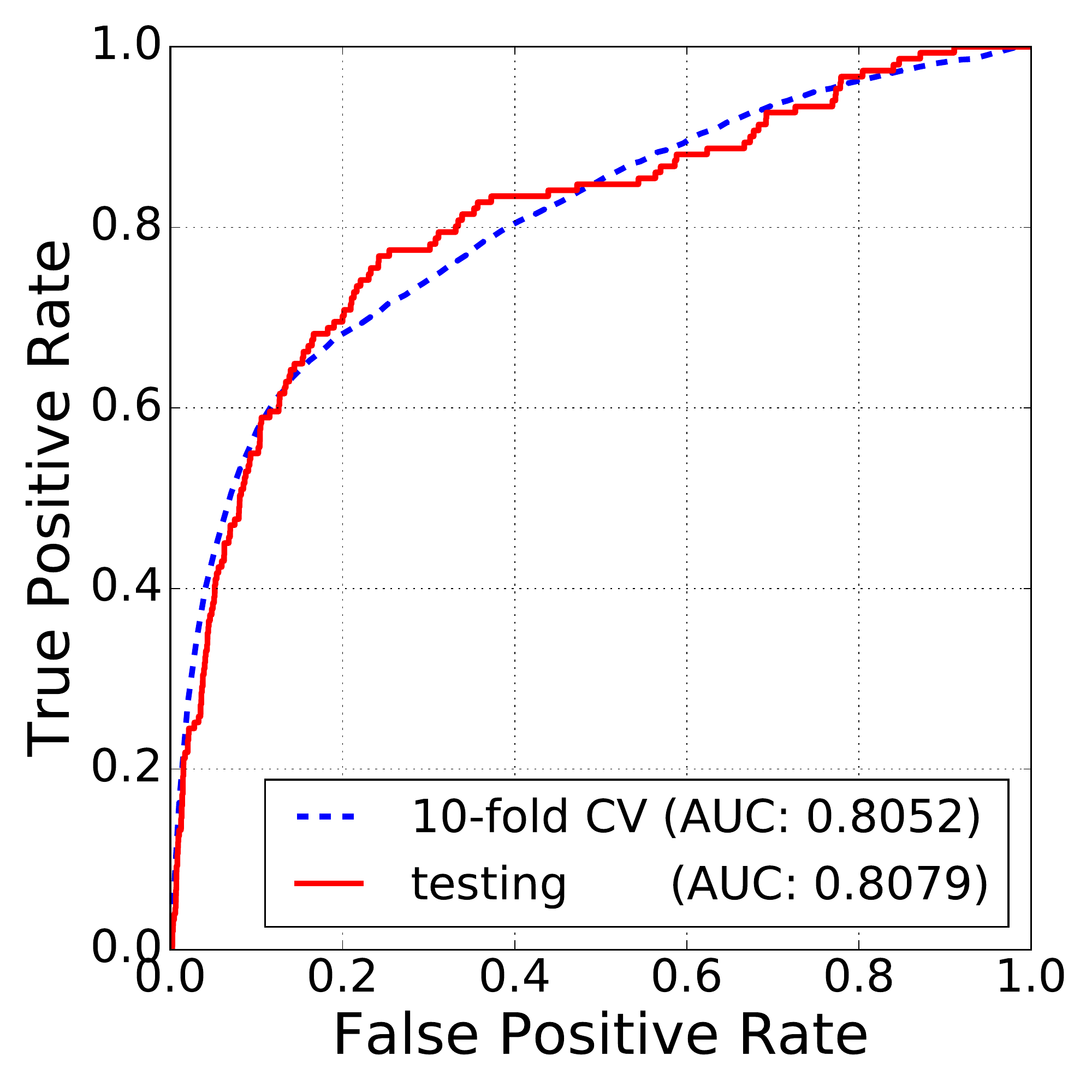}
        \caption{SVM}
        \label{fig:svm_curve}
     \end{subfigure}
          \caption{ROC curves of Random Forest and SVM}\label{fig:animals}
          \label{fig:roc_curve}
\end{figure}

The ROC curves for the training and testing performances are shown in Figure~\ref{fig:roc_curve}. 
All the results are averaged over 10 trials. The most important metric in this case is the true positive rate (TPR), i.e., how many fires were correctly predicted as positive in our model. 
The SVM model was able to predict $71.36\%$ of the fires in 2014-2015, at a false positive rate (FPR) of $20\%$, 
which was deemed practically useful for AFRD --- potential to save lives (by achieving a higher TPR) significantly outweighs the increase in FPR. At the same time, a high FPR facilitates more inspections of risky buildings, which is also beneficial. In practice, AFRD can adjust the TPR/FPR ratio to match their risk aversity and inspection capacity.
The Random Forest model achieved a slightly lower TPR of $69.28\%$ at the same FPR, but had a higher area under the ROC curve (AUC). 
Considering how few fires occur (only about 6\% of the properties in our total dataset had fires), these results are much more predictive than guessing by chance.

False positives (FPs) provide important information to AFRD. As our testing period was the final year in our dataset, 
it is possible that some of those FP properties may actually catch fire in the near future. These properties share many characteristics with those  that did catch fire, and should likely be inspected by AFRD.

\subsection{Further Discussion of the Models}
In this section, we discuss some insight we obtained while conducting the experiments. First, there is a mismatch between the meaning of labels in the training and testing datasets. The training labels represent fires that happened in a relatively long period of time, whereas the testing labels represent fires in a single year. One way to address this issue would be to expand each properties into multiple examples, one for each year. Each example is then a properties for a particular year, and the corresponding label indicates whether there was a fire in that year. Using this approach, however, did not improve the performance in our experiments. The reason is that most of our variables are static, such as floor size and zip code, and only a few variables are time-dependent, such as the age of the building and the time since last inspection. Therefore, expanding the properties only gives us many similar examples. However, this approach would potentially be helpful after collecting other dynamic information in the future, such as violations of health codes, sanitation ordinances, or other information from relevant city agencies.

Another important issue is whether the performance of predicting fires is consistent in different testing time periods. 
To test this, we tried different time windows for training, and for each window, we evaluated its prediction performance for the subsequent year. 
For each time window, we repeated the process described in Section~\ref{sec:evaluate}, including grid search and cross validation, and finally used the best model to predict fires  in the following year. 
The results are shown in Table~\ref{table:split}. The performances decrease slightly for shorter training periods. This is due to fewer positive training examples, especially in the period of 2011-2012, which only consists of eight months of data (July 2011 to March 2012). However, this is still significantly better than guessing by chance, which demonstrates that we were not just ``lucky" in predicting fires for a particular year.

Finally, it is helpful for us and for AFRD to know which features are the most effective predictors. The Random Forest model presents a natural way to evaluate feature importance: for each decision tree in the Random Forest, the importance of a feature is calculated by the ratio of examples split by it. 
The final importance is then averaged among all trees. 
The top ten most predictive features are displayed in Table~\ref{table:top_features}.
Collectively, they capture the intuitive insight that buildings of a larger size or those containing more units (thus more people) would have higher probability of catching fire. 

We also tried logistic regression, a linear model, to estimate each feature's importance based on the corresponding weight coefficient in the model. We found that the top features in the logistic regression were very different from the ones in Random Forest. All were binary features indicating either a particular neighborhood or property owner. Some neighborhoods have either very high or low fire rates, and logistic regression tends to assign large positive or negative weights to them, respectively. However, since each of these features is only good at predicting a small number of properties within a certain area but does not predict well on the overall data, they are not chosen in the first few iterations of a decision tree.

\begin{table}
\small
\sffamily
\centering
\begin{tabular}{ccc}
\toprule
& \multicolumn{2}{c}{Testing AUC of the following year} \\
\cmidrule{2-3} 
Training window    & Random Forest & SVM  \\ 
\midrule
2011-2012       & 0.7624   & 0.7614       \\  
2011-2013       & 0.8030   & 0.7914        \\  
2011-2014       & 0.8246   & 0.8079        \\  
\bottomrule
\end{tabular}
\caption{Testing AUC of each year}
\label{table:split}
\end{table}

\begin{table}
\small
\sffamily
\centering
\begin{tabular}{ll}
\toprule
  & Top 10 features  \\
\midrule
1   & floor size     \\
2   & land area       \\
3   & number of units          \\
4   & appraised value            \\
5   & number of buildings              \\
6   & total taxes             \\
7   & property type is multi-family             \\
8   & lot size             \\
9   & number of living units             \\
10  & percent leased             \\
\bottomrule
\end{tabular}
\caption{Top-10 features in Random Forest}
\label{table:top_features}
\end{table}

\subsection{Assignment of Risk Scores}
After we built the predictive model, we then applied the fire risk scores of each property to the list of current and potential inspectable properties, so that AFRD could focus on inspecting the properties most at risk of fire. To do this, we first computed the raw output of our predictive model for the list of  properties we used to train and test the model. 
This generated a score between 0 and 1, which we then mapped to the discrete range of 1 to 10 that is easier for our AFRD colleagues to work with. 
Then, based on visual examination of the clustering of risk scores, we categorized the scores into low risk (1), medium risk (2-5), and high risk (6-10). These risk categorizations were intended to assign a manageable amount of medium risk ($N = 402$) and high risk properties ($N = 69$) for AFRD to prioritize.  

We then needed to find out which of the  properties with risk scores were in the lists of 2,573 current annually inspected properties and 6,096 potentially inspectable properties. Because of the lack of a consistent property ID across the various datasets used to develop the risk model, the currently inspected and potentially inspectable properties were spatially joined with the properties in the risk model, based on their geo-coordinates or addresses. After joining, we were able to assign risk scores to 5,022 of the 8,669 total commercial properties on the inspection list (both currently inspected [2,573] and potentially inspectable [6,096]).

\section{Impact On AFRD and Atlanta}

\subsection{Previous Inspection Process}
Our goal in developing the Firebird framework was to help the Atlanta Fire Rescue Department (AFRD) and other municipal fire departments improve their identification and prioritization of commercial property inspections. Before considering the impact our work had on that process, it is important to first describe the previous process of commercial fire inspections in Atlanta. First, fire inspectors at AFRD received a list of properties to inspect every month, which had been inspected during that same month in the previous year. The existing process for adding new commercial properties to the list of required inspections was extremely ad hoc, without a formal notification process from other city departments when new buildings were built or occupied, or new businesses registered. It was largely the responsibility of individual fire inspectors to notice new inspectable properties and initiate an inspection process while driving to another inspection site. Moreover, there was no formal process used by the inspectors to prioritize their monthly inspections based on risk, or even to schedule their daily inspections based on proximity to other inspections. In other words, it is very possible that an inspector could return to the same business complex multiple times throughout the year conducting inspections on adjacent properties, which is not the most effective use of municipal resources. In addition, at present, the City of Atlanta Code of Ordinances does not specify the frequency of inspections based upon risk or other factors. As a result, inspections are effectively binary; regardless of potential fire risk, a property either gets an annual inspection in the same month every year, or it is unlikely to be inspected at all.

\begin{figure*}
\centering
    \includegraphics[width=0.86\textwidth]{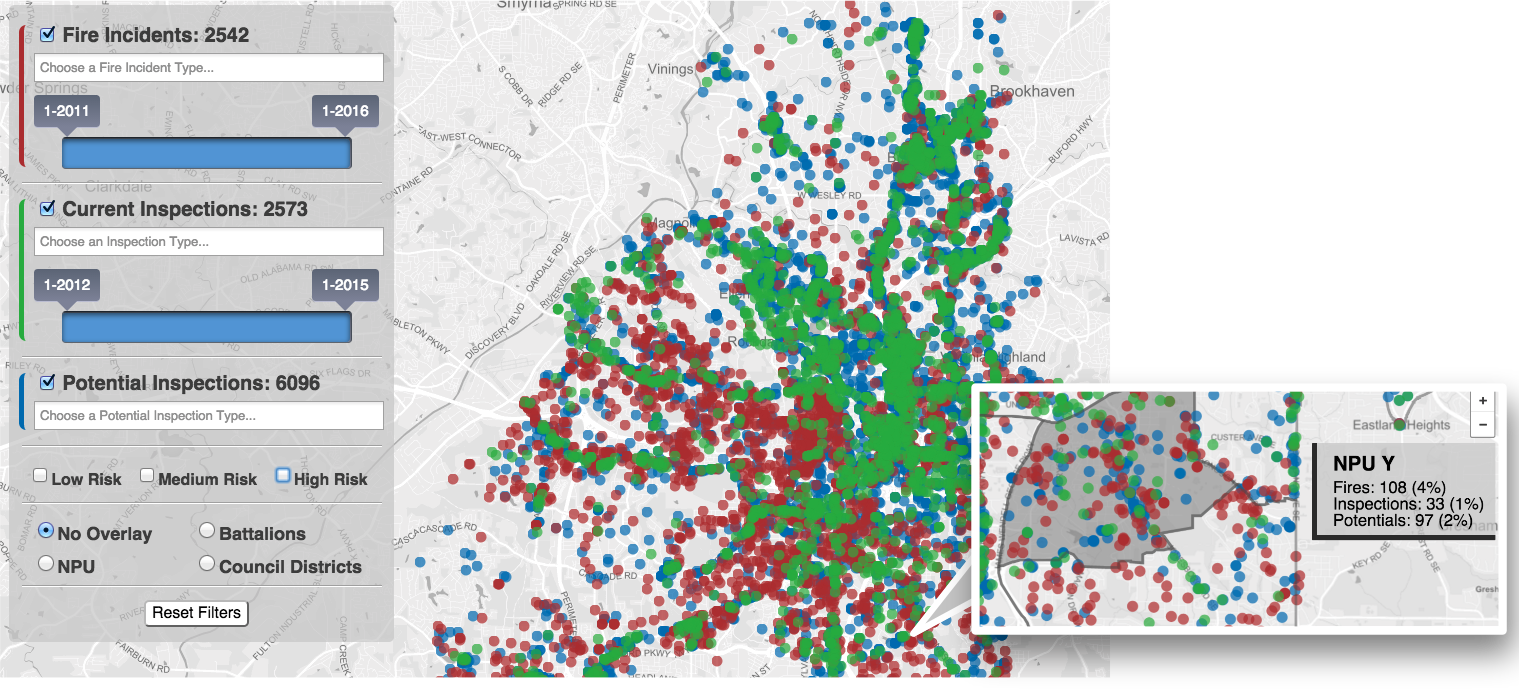}
\caption{\textbf{Interactive map of fires and inspections.} 
The colored circles on the map represent fire incidents, currently inspected properties, and potentially inspectable properties in red, green, and blue, respectively. 
Inspectors can filter the displayed properties based on property usage type, date of fire or inspection, and fire risk score.
\textit{Callout:} activating the Neighborhood Planning Unit overlay allows an inspector to mouse-over a political subdivision of the city to view its aggregate and percentage of the fires, inspections, and potential inspections.
}
\label{fig:map}
\end{figure*}

\subsection{Technology Transfer to AFRD}
After developing the Firebird framework, we first provided AFRD's executive staff and Community Risk Reduction Section with a dataset of all commercial properties in Atlanta that fit their criteria for inspection. This included a shorter list of 6,096 new inspectable properties which are of the top 100 currently inspected property usage types (e.g., restaurants, motor vehicle repair facilities, etc), and a longer list of all commercial properties (19,397) that fit any property usage type that had been previously inspected. This dataset included the associated building- and parcel-level information for those properties in the form of a CSV file, with a subset of those properties (5,022) assigned a fire risk score. We then provided AFRD with an interactive map-based visualization tool, as part of the Firebird framework, for the fire inspectors and AFRD executive staff to use to augment their policy and decision-making processes. The map in Figure \ref{fig:map} was made using the open source map-making tools Mapbox and Leaflet to create the base map layer. Then, using the Javascript visualization library D3.js, we displayed differently colored circles on the map to represent fire incidents, currently inspected properties, and potentially inspectable properties in red, green, and blue, respectively, using their longitude and latitude coordinates. 

We also built a user interface for the Firebird map developed through discussions with the AFRD Assessment and Planning Section, and refined by incorporating feedback from fire inspectors and AFRD executive staff. The map includes an information panel for displaying property information when hovering over a property on the map, such as its business name, address, occupancy usage type, date since fire incident or inspection, and fire risk score, if available. The map also includes a user interface panel with the ability to filter the fire incidents, the currently inspected, and the potentially inspectable properties according to their property usage type, the date of fire incident or inspection, and their risk score. Finally, we incorporated a set of regional overlays requested by the AFRD executive staff, including the AFRD battalions, the Atlanta Neighborhood Planning Units (NPU), and Council Districts, of which the latter two are political subdivisions of the city. We included dynamically updated counts and percentages for currently displayed fire incidents, current inspections, and potential inspections for each of those regional overlays (shown in Figure 4), to be displayed when hovering over them, so the AFRD executive staff could make decisions at a battalion, NPU, or Council District level. This map, and the Firebird framework in general, could be used as a powerful tool for supporting data-driven conversations about personnel and resource allocation and inspection decisions, and may even be used to inform decisions regarding community education programs for fire safety and prevention.

\subsection{Impact on AFRD Processes}
After receiving the dataset of properties needing inspection, prioritized according to their fire risk score, AFRD has begun integrating the results of the analytics into their fire inspection process. Increasing the number of annual inspections by 6,096 (237\%) overnight was not feasible without significant changes in organizational processes, local ordinances, or increased staffing. As an initial effort, AFRD assigned the 69 high-risk properties to the inspectors covering those respective areas. Of those, 27 had current or out-of-date fire safety permits that required re-inspection, 13 properties required new permits, and 15 properties recently went out of business. The remaining properties were found to not require a fire safety permit. Most significantly, the inspectors assigned to review these properties found a total of 48 violations that needed to be addressed to meet the fire code. As AFRD continues working through the list of potentially inspectable properties in descending order of risk, the sheer number of additional inspections (increased workload) and the number of potential violations identified and mitigated (positive outcomes) has already begun to have a transformative impact on the daily operations of AFRD's fire inspection process.

In addition to the immediate impact on the daily property inspections, the results of this work have stimulated important conversations within the executive leadership of AFRD and the Assessment and Planning Section about 1) how to more effectively allocate inspection personnel; 2) how to update and utilize the model to provide dynamic risk data in real time (e.g., on a monthly basis when new inspection assignments are given to the inspectors); 3) how to motivate increased data sharing between various government departments such as the Office of Buildings and AFRD; 4) how to give teams of firefighters access to fire safety permit and violation information when they respond to a fire emergency at that commercial property; and 5) how to extend the risk prioritization to residential properties using more behavioral data such as noise or sanitation ordinance violations, and consumer data from companies like Experian or ESRI.  

Though there are many more inspectable properties than AFRD currently has the personnel capacity to handle, AFRD has already begun to take steps toward a more efficient use of their existing personnel, by discussing how to assign inspectors to regions with a higher proportion of properties requiring inspection, rather than by the geographical assignment to fire battalions currently in use. In addition, they have begun to discuss altering properties' inspection frequencies to reflect their fire risk levels. By prioritizing future inspections and more efficiently allocating inspection personnel to target the commercial properties most at risk of fire, we hope this work will lead to a reduction in the frequency and severity of fire incidents in Atlanta. We also hope that this framework can be instructive for other municipal fire departments to improve their fire inspection processes.

\section{Challenges}
As fire departments in many municipalities embark on more data-driven fire risk inspection policies and practices, several challenges we encountered could prove instructive for others. As with many governmental data science initiatives, the practical application of the predictive model has been contingent on local politics, organizational inertia, and existing policies, or what \cite{carr05} calls the ``politics of place.' While AFRD was an active and engaged partner throughout this project, securing access to clean and usable data proved a challenge. At the time of this writing, the City of Atlanta's Office of Buildings, Office of Housing, and AFRD do not share a unified database of buildings or a shared building identification numbering convention, and thus, the process of joining various datasets was more technologically difficult than it might otherwise have been. Even the seemingly simple task of discovering which properties in the Office of Building's dataset were also in the AFRD dataset required a rather elaborate process of fuzzy text-matching and address verification. In addition, our ability to leverage regularly updated dynamic data was similarly hindered by the difficulty of data sharing among city departments. For instance, the Office of Building's Business License database may very well be updated regularly, but without a pipeline in place for those updated data to be used by AFRD, businesses may close without AFRD knowing, causing inspectors to waste time attempting to inspect closed businesses. These challenges could be mitigated if each department with a vested interest in municipal commercial properties and structures worked more closely to share their data and information. 

Entrenched organizational processes may similarly hinder the adoption of new methods of identifying properties to inspect. While, in theory, the fire inspection process targets properties that are required by city ordinance to have a fire safety permit (e.g., restaurants over a certain capacity, auto repair facilities, etc.), in practice there is no existing organizational procedure at AFRD and in many other cities to systematically add properties to the list of regular inspections, or to determine their frequency of inspection \cite{garis14}. In this work, while we have created and employed an innovative method of identifying new properties to inspect, until inter-departmental data-sharing becomes widespread, this process would need to be redone on a regular basis as new businesses open, close, or change usage type. Further, though we created an interactive map for visualization of various types of property inspections, such a tool has not previously been part of AFRD fire inspectors' regular workflow, and this novelty presents a barrier for adoption, as seen in \cite{dobrica12}. Finally, with the number of inspectable properties increasing by up to 237\%, there is no clear incentive for the fire inspectors to work more efficiently to increase their individual number of property inspections per month.

From a policy standpoint, the existing municipal fire code in Atlanta requires that inspections occur regularly for a specified set of commercial property types, but it is not clear that those property types require inspection with equal priority or frequency. After using the results of this work for determining individual property fire risk, the AFRD should begin a conversation about how best to revise the municipal fire code to reflect differences in inspection priority and frequency due to the fire risk associated with various property types. Prior work in \cite{garis14} has similarly suggested revisions of the British Columbian fire code from being a reactive, inflexible document based on tradition and intuition to a data-driven, responsive, and pro-active document that incorporates information about fire risk. Finally, AFRD's policies for property fire inspections are primarily geared towards commercial properties, yet the majority of the fires in Atlanta occur in residential properties. This will require additional rethinking of their residential community fire safety and prevention education programs.
 
As in many municipalities, fully leveraging the power of analytics to improve fire safety in Atlanta will require a significant rethinking of how to approach and manage city operations such as fire inspections, and how to best facilitate data sharing practices between different city agencies.

\section{Conclusions and Future Work}
Due to the large number of commercial properties in Atlanta potentially requiring inspection and the limited inspection personnel capacity of the Atlanta Fire Rescue Department (AFRD), as in many other municipalities, there is a need for a data-driven prioritization of commercial property inspections. In this paper, we provide the Firebird framework: a re-usable method for municipal fire departments to identify and prioritize their commercial property fire inspections based on each property's fire risk. Our work first provides a clear process for joining disparate data sources from multiple municipal departments and private sources to identify new inspectable properties based on currently inspected property types. We were able to identify 6,096 new inspectable properties, comprised of the top 100 property types currently inspected by the AFRD, and a total of 19,397 new inspectable properties comprised of all currently inspected property types by AFRD. 

We next present a method for predicting fire risk for each commercial property. Our models used 58 building- and parcel-level variables to predict fires in 8,223 properties, 5,022 of which are on the list of properties requiring inspection. Specifically, we trained SVM and Random Forest models using data from 2011-2014 to predict fires in 2015. 
At a false positive rate of 20\%, the SVM and Random Forest models were able to predict 71.36\% and 69.28\% of the fires in that year, respectively. Furthermore, even the false positives provided valuable insight, since they represent properties with high risk of catching fire, that likely should be inspected by AFRD. We also identified features that are highly related to fires. From the Random Forest model, we learned that features related to building size, number or units, and value were most predictive. On the other hand, the logistic regression model revealed certain neighborhoods and property owners that associate with very high or low fire rates. We then converted these results to a risk score for each property, and were able to apply these scores to 5,022 currently inspected and potentially inspectable properties (1,975 currently annually inspected and 3,047 potentially inspectable), with 454 of those properties having a medium or high risk score (188 currently inspected and 266 potentially inspectable properties). Finally, we incorporated those scores into our joined dataset of property inspections and visualized each of the properties on an interactive map, with their associated property information and risk score, for use by AFRD to augment their decision-making and inspection processes. 

\textbf{Research Directions.} Future research should seek to refine, expand, and further validate our prediction model. Due to missing or erroneous entries in the data sources, we were only able to incorporate 8,223 properties into our predictive fire risk model, out of the more than 20,000 commercial properties in the city. In addition, because of the lack of integration across city department datasets and a lack of completeness in many of our datasets, we were only able to provide risk scores for 5,022 of the 8,669 current and potentially inspectable properties in the dataset we provided to AFRD. Researchers working with other municipal fire departments might train their models on a dataset that has fewer building- or parcel-level information variables, but may be applicable to more properties. Other work could improve the accuracy of the model by incorporating additional dynamic sources of data, such as violations of prior fire inspections, data from the Department of Health and Wellness inspections, information from the Certificates of Occupancy, or other, more behavioral sources, such as sanitation or noise violations, as seen in \cite{copeland15}, rather than the largely static building- and parcel-level data that we used. In addition, more research needs to be done on the usefulness and usability of an interactive map to display inspection, and how the inspectors or executive staff of a fire department could use it in different ways to inform their day-to-day planning, decisions, and operations. One step that municipal government agencies can take towards implementing this framework is to generate a unique Building Identification Number (BIN), used by all stakeholders, such as the Office of Buildings, Office of Housing or city planning departments, as well as the Fire and Police Departments. This would allow for an easier joining of various disparate sources of data, without the need for the address validation, text matching, and other complex and potentially error-generating processes for joining datasets.

Identification, selection, and prioritization of risky properties for fire inspection can be very difficult for cities that do not have an integrated data platform, because some municipal agencies may have relevant property and structure information that is isolated from other local data sources, and which may not have a regular, timely process for updating. Our work can be a model for improving the complex process of property inspection identification and prioritization. We hope that other fire departments and other municipal organizations interested in applying a data-driven analytic approach to their property inspections can use the framework outlined here for identifying and prioritizing inspections based on risk criteria. Additionally, our experience joining isolated datasets from different government departments, commercial data, and open data sources could be invaluable for many cities that want to begin utilizing data science for a smarter city, without requiring a significant financial investment. We hope the impact from our work may further promote the beneficial use of open public sector data, both in the city of Atlanta and elsewhere.


\section{Acknowledgments}
We want to thank our partners at the Atlanta Fire Rescue Department for their support and willingness to re-examine their existing processes, especially Chief Baker, Chief Rhodes, and the staff of the Community Risk Reduction Section. 
We also want to thank the Georgia Institute of Technology and the Data Science for Social Good program for their funding and academic support for this research.

%
\bibliographystyle{abbrv}
\bibliography{sigproc}

\balancecolumns 
\end{document}